# A 5.5 ps Time-interval RMS Precision Time-to-Digital Convertor Implemented in Intel Arria 10 FPGA

Jie Kuang, Yonggang Wang

*Abstract*—As an important part of the field programmable gate array (FPGA) market, Intel FPGA has also great potential for implementation of time-to-digital convertor (TDC). In this paper, the basic tapped delay line (TDL) TDC structure is adapted in Intel Arria 10 FPGA, which is manufactured with 20 nm process technology. Because of the serious bubble problem for FPGA made by state-of-art process, the ones counter encoding scheme is employed to maintain the delay elements in TDL resolvable for achieving high TDC time precision. The test of TDC bin width reveals that the characteristics of the delay chain are highly consistent with the fundamental structure of logic resource in the FPGA. To improve TDC time precision, four TDLs are combined parallel for final TDC implementation. Using two identical TDC channels, the average RMS precision for measurements of time-intervals in the range from 0 to 50 ns reaches 5.45 ps. The test results demonstrate that high performance TDC can be implemented in current Intel main-stream FPGAs as well.

## I. INTRODUCTION

IMPLEMENTATION of time-to-digital convertor (TDC) based on field programmable gate array (FPGA) is widely investigated in the past decades[1]-[3]. Except of several designs in early stage using Intel low-end FPGAs [3], most current designs are being implemented with Xilinx FPGA chips, in particular using FPGAs made with 28 nm and more advanced process [4][5]. One possible reason for the infrequent use of Intel FPGA is that there are very few bottom-level design features opened by Intel design software, so that the fundamental logic resource of FPGA is less controllable which may reduce the designer's confidence. However, as an important part of the FPGA market, Intel FPGA has many powerful aspects that other FPGA cannot replace, which makes them used in wide range of applications, including time-measurement domains. Design and Integrating TDC inside of Intel FPGAs is the same important and necessary.

Manuscript received January 30, 2018. This work was supported by the National Natural Science Foundation of China (NSFC) under Grants 11475168 and 11735013.
Authors are with the Department of Modern Physics, University of Science and Technology of China, Hefei, Anhui 230026, China. Corresponding author: wangyg@ustc.edu.cn.

Nowadays, FPGAs manufactured with 28 nm and more advanced process technology are very common. They are becoming more and more powerful and cost-effective digital circuit and system design platforms. Because both Xilinx FPGA and Intel FPGA have pre-designed hardware adders, which carry logic can be cascaded to construct specific TDLs, both FPGAs are feasible for TDL style TDC implementation. According to our experience of TDC in Xilinx FPGA [4][5], the TDL constructed in FPGAs made with 28 nm and more advanced process has very serious bubble problem. If these bubbles are simply discarded by the following thermometer-to-binary encoder, the bubble delay elements will be folded into its neighbors resulting in a decrease of effective TDC bins, which is negative for high precision TDC implementation. To make all the delay elements in TDL resolvable, in our previous work, we have proposed two effective solutions: one is the bin realignment method [4] and the other is ones counter encoding method [5]. Since ones counter encoding method has advantages of robustness and simplicity of implementation, it will be applied in this paper to the TDC design using Intel Arria 10 FPGA that is made with 20 nm process. Because the fundamental structures of pre-designed logic resource in both FPGAs are rather different, it is expected that the TDL properties in both FPGAs are different, so that the achievable TDC performance should be also different. We will first measure the TDL property of the Intel FPGA, and then improve it by a parallel multi-chain structure. A practical 5.45 ps RMS precision TDC with 250 M event per second measurement throughput at a system clock rate of 500 MHz is implemented. The test results in this paper are comparable with that obtained in our previous work [5] with the same TDC structure but using the counterparts in Xilinx FPGA. It is significant to understand their difference from the perspective of their underlying structural features.

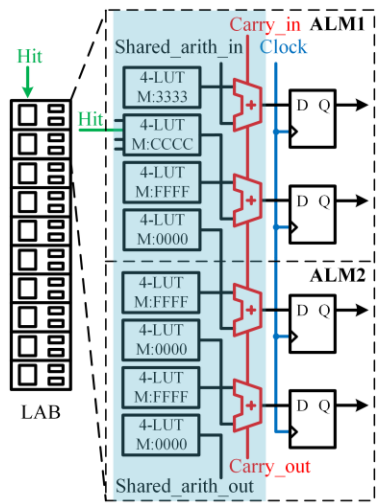

Fig. 1.  Carry logic in a LAB is cascaded into a part of delay line

## II. TDC Implementation

FPGA based TDC is mainly using a coarse counter that runs at system clock rate to give out a coarse timestamp and a TDL for time interpolation to yield a sub-clock period resolution. In Intel FPGA, adaptive logic modules (ALM) are fundamental logic resource units, which have two dedicated carry logic. Ten ALMs together as a group is called as a logic array block (LAB or MLAB). Cascading them in one LAB form a carry chain with 20 taps. The whole TDL is constructed by cascading fast carry chains in multiple LABs. The TDL status will be latched out by their accompanied Flip-Flops as shown in Fig. 1. With a hit signal propagating along the TDL entering through a LUT logic, the recorded TDL status is converted to a binary code by ones counter encoder and then will be calibrated by an online calibration table. The combination of the calibrated fine timestamp and the coarse timestamp is final measurement. We will measure the TDL property of the Intel FPGA by constructing a TDC with a single TDL.

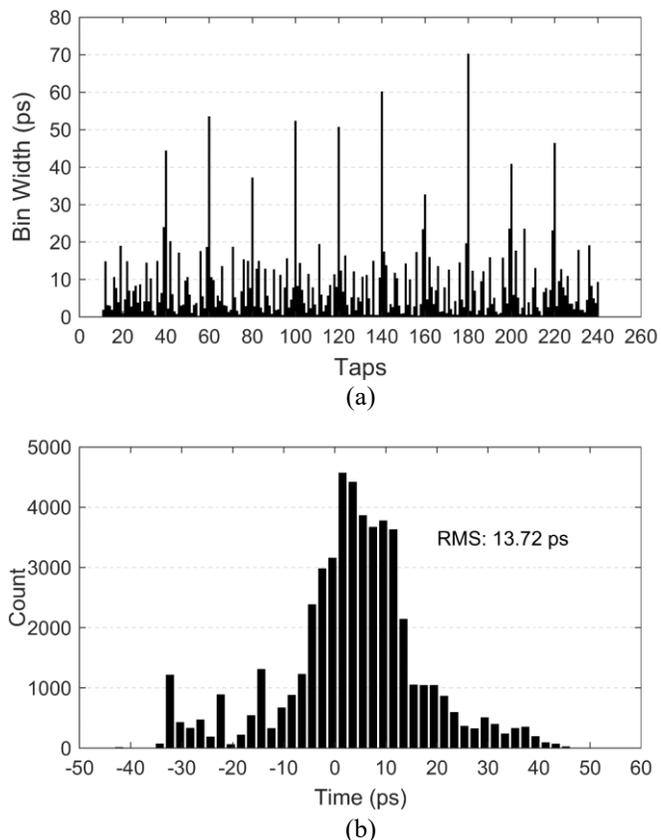

Fig. 3. (a) Bin widths and (b) test histogram using two single-TDL TDC channels to measure a time-interval of about 3.5 ps using calibration.

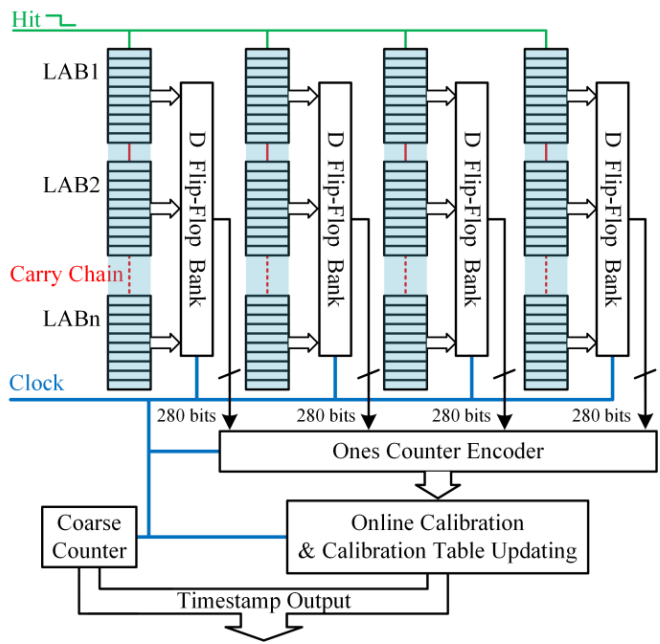

Fig. 2. Four TDLs merged together to construct a high precision TDC.

To improve the time precision further, we combine four TDLs parallel as one merged TDL-TDC as shown in Fig. 2. In the 4-chain architecture, every chain has 280 delay elements to ensure the total delay time of the merged TDL is larger than the system clock period.

## III. Test Results

### A. Single-TDL TDC test

We first implemented two identical single-TDL TDCs on the FPGA (10AS066N3F40E2SGE2) in an Arria 10 SoC development board. Fig. 3(a) shows the measured TDC bin widths, and Fig. 3(b) is the typical test histogram when we used them to measure the time interval about 3.5 ps. It is obvious that there is a special big TDC bin in every 20 bins, i.e., the $40^{th}$, $60^{th}$ bins and so on. This TDL property is consistent with the structure of TDL. These ultra-wide periodic bins are the positions where the hit signal propagates cross LAB boundaries. With these big bins, the time-interval RMS precision is measured as 13.72 ps, which is worse than the TDCs we implemented using Xilinx Kintex-7 FPGA [4][5].

### B. Multi-TDL TDC test

As the work we did using a Kintex-7 FPGA [5], two identical TDC channels using 4-chain merged TDLs followed by the ones counter encoder were implemented and evaluated. The TDC bin widths were measured by the code density test as shown in Fig. 4(a) and the derived bin width histogram is shown in Fig. 4(b).

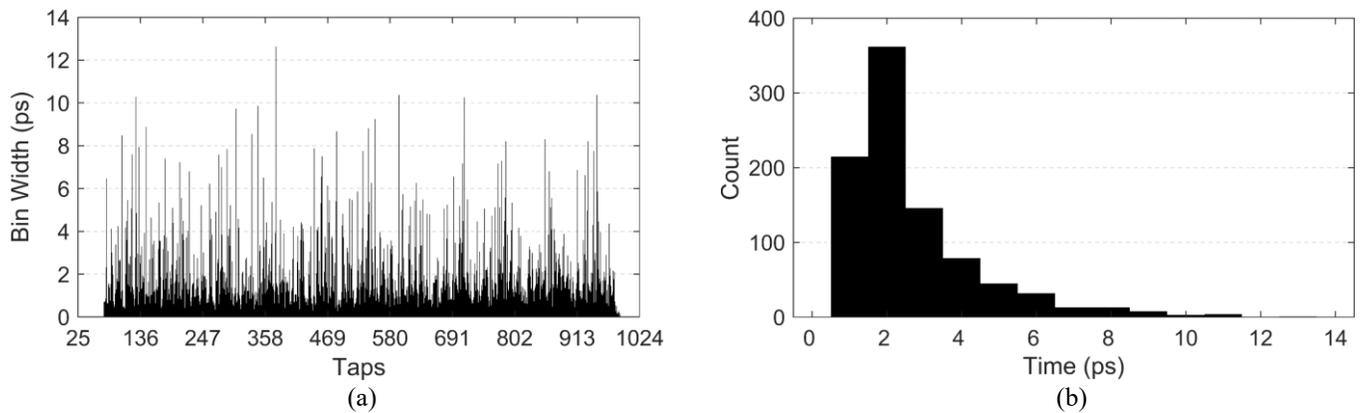

Fig. 4. (a) Bin width and (b) bin width histogram of a 4-chain merged TDL TDC channel.

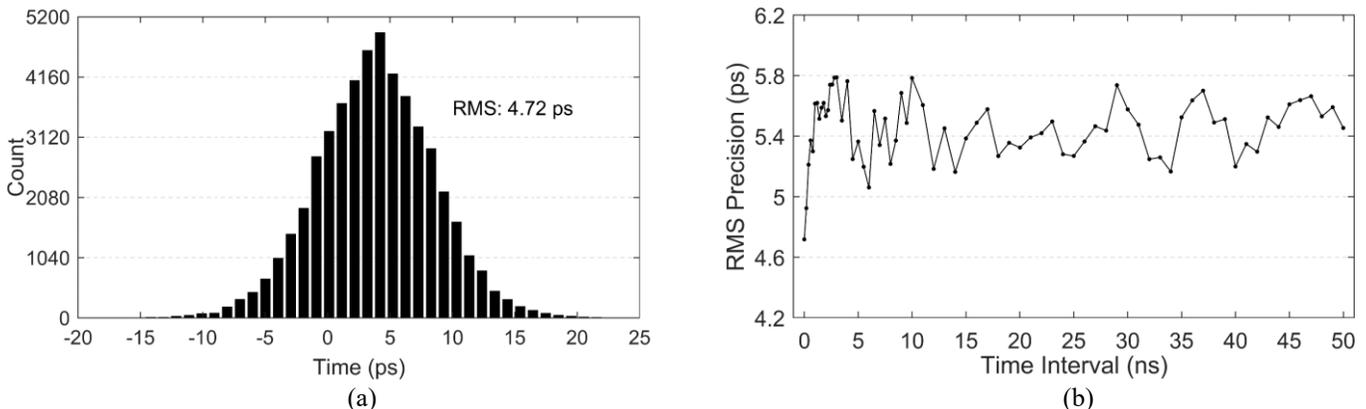

Fig. 5. (a) Measurement histogram of time-interval of 3.83 ps. (b) RMS precision varying with time-intervals from 0 ns to 50 ns.

We can see that the ultra-wide bins disappeared. The total number of effective bins is 921 in the range from bin 71 to bin 991, which interpolates one system clock period (2 ns), so that the average bin width can be calculated as 1LSB = 2000/921 = 2.17 ps. Fig. 5(a) is a typical shape of test histogram for a time-interval measured by the two TDC channels. The RMS precision for time-intervals can be calculated from their histograms. The RMS precision varying with time intervals in the range from 0 ns to 50 ns are measured as shown in Fig. 5(b). The average value of time precision over the whole measurement range is 5.45 ps. The test results show that the granularity and uniformity of multi-chain TDC bins are substantially improved, and the TDC performance improvement is significant.

## IV. CONCLUSION

Using the structure of four TDLs parallel, a practical 5.45 ps time-interval RMS precision TDC with 250 M measurement throughput is achieved in Intel Arria 10 FPGA. This work demonstrates that both Intel and Xilinx FPGA are suitable for high performance TDC implementation. The measured result of TDL property is highly consistent with its underlying structure features, which indicates Intel FPGA has slightly poor ability for achieving TDC. Although the emerged TDL has very similar property with that we measured using Xilinx Kintex-7 FPGA, which is made with 28 nm process, the implemented TDC time performance is still poorer than the 3.9 ps RMS precision achieved in the Xilinx FPGA. One reason is likely to be that the hit signal leading to the TDL must go pass a LUT logic. The extra jitter of the LUT logic spoils the final achievable time precision.